\documentclass[sigconf]{acmart}

\AtBeginDocument{%
  }
\usepackage{tcolorbox}
\usepackage{algorithm}
\usepackage{algpseudocode}
\usepackage{enumitem}
\usepackage{multirow}

\definecolor{DarkGreen}{RGB}{0,128,0}
\definecolor{BrightRed}{RGB}{220,0,0}
\newcommand{\gain}[1]{{\textbf{\textcolor{DarkGreen}{(#1\%)}}}}
\newcommand{\loss}[1]{{\textcolor{red}{(#1\%)}}}
\begin{document}

\title{EpiDroid: Dependency-Guided Recomposition for Deep State Discovery in Mobile GUI Testing}

\author{Jiahui Song}
\email{songjah@zju.edu.cn}
\affiliation{%
  \institution{Zhejiang University}
  \city{Hangzhou}
  \country{China}
}

\author{Jiaxin Zhi}
\email{2023091602025@std.uestc.edu.cn}
\affiliation{%
  \institution{University of Electronic Science and Technology of China}
  \city{Chengdu}
  \country{China}
}

\author{Kangjia Zhao}
\email{konkaz@zju.edu.cn}
\affiliation{%
  \institution{Zhejiang University}
  \city{Hangzhou}
  \country{China}
}

\author{Chen Zhi}
\authornote{Corresponding author.}
\email{zjuzhichen@zju.edu.cn}
\affiliation{%
  \institution{Zhejiang University}
  \city{Hangzhou}
  \country{China}
}

\author{Junxiao Han}
\email{hanjx@hzcu.edu.cn}
\affiliation{%
  \institution{Hangzhou City University}
  \city{Hangzhou}
  \country{China}
}

\author{Xinkui Zhao}
\email{zhaoxinkui@zju.edu.cn}
\affiliation{%
  \institution{Zhejiang University}
  \city{Hangzhou}
  \country{China}
}

\author{Nan Wang}
\email{wangnan02026@163.com}
\affiliation{%
  \institution{Shenzhou Aerospace Software Technology Company Limited}
  \city{Beijing}
  \country{China}
}

\author{Shuiguang Deng}
\email{dengsg@zju.edu.cn}
\affiliation{%
  \institution{Zhejiang University}
  \city{Hangzhou}
  \country{China}
}

\author{Jianwei Yin}
\email{zjuyjw@cs.zju.edu.cn}
\affiliation{%
  \institution{Zhejiang University}
  \city{Hangzhou}
  \country{China}
}

\begin{abstract}
 The increasing scale and complexity of mobile applications make automated GUI exploration essential for software quality assurance. However, existing methods often neglect state dependencies between test fragments, which leads to redundant exploration and prevents access to deep application states. We introduce \textbf{EpiDroid}, a black-box, pluggable framework that augments existing explorers through semantic state dependency awareness. EpiDroid distills raw traces into stable test fragments to extract underlying dependencies. It then employs a Recomposition-Replay paradigm to perform impact reasoning via LLM and deterministic replay on high-value mutable state elements. Through iterative feedback, EpiDroid refines the state-dependency graph to systematically reach deep application states. We integrated EpiDroid into both industrial and state-of-the-art research tools and evaluated it on 20 real-world apps. The results show that EpiDroid consistently improves the performance of all baselines, increasing average code coverage by 10--28\% and delivering 3--4$\times$ more coverage gain compared to continuing the baselines alone from the same starting point. This demonstrates that dependency-guided recomposition unlocks deep states that forward exploration cannot access, irrespective of additional budget.
\end{abstract}




\maketitle

\section{Introduction}
\label{sec:intro}

Mobile applications are now fundamental to daily life, spanning domains from digital payments to social networking and navigation. As mobile apps grow in scale, feature richness, and interaction complexity, testers increasingly rely on automated GUI testing tools to exercise large state spaces within limited time budgets~\cite{Li23b}. However, deep exploration remains a persistent bottleneck. Prior reviews show that mobile GUI testing is still fundamentally constrained by effectiveness, efficiency, and practicality, typically measured through coverage, bug detection, and execution cost or time~\cite{Arn18,Sai20,Su21}. In practice, many important states remain difficult to reach because they depend on specific prior interactions, data mutations, account context, or configuration choices that are not exposed by shallow forward exploration.

We observe that this bottleneck often stems from a limited understanding of how local behaviors compose into global ones. Many states cannot be reached by simply deepening a single execution path; instead, they emerge from the synergy of multiple interaction snippets under specific semantic conditions. For instance, certain features may only surface when specific configurations are enabled or particular data objects exist. While these states are semantically reachable, their triggering conditions are often fragmented across different execution paths and temporal stages.

This limitation persists across traditional and recent methods. Prior work improved Android GUI testing scalability through better event generation, search, and UI modeling, and through stochastic or learning-based guidance~\cite{Mac13,Mao16,Su17,Pan20,Rom21}. Recent LLM-based approaches enhance the semantic quality of local exploration decisions~\cite{Liu23,Liu23c,Yoo24,Wan25,Wen23c}. Despite these advances, existing methods remain constrained by a forward-execution paradigm: they optimize what to do next along the current trajectory but lack explicit abstractions to reason about how behaviors across execution branches relate and recombine to satisfy hidden dependencies. Consequently, these tools often spend substantial effort on redundant shallow exploration while failing to reach deep, semantically conditioned states.

Existing studies also highlight this missing capability. They show dependency information can improve exploration by exposing relations among UI elements and system events \cite{Guo20}. Others demonstrate that complex functionality often emerges from combinations of short interaction sequences \cite{Wan20}, that richer launching contexts can reveal distant activities \cite{Yan20}, and that replaying from discovered states can be more effective than reconstructing them from scratch~\cite{Don20,Guo19}. Collectively, these results point to a common insight: deep-state reachability depends not only on better next-action selection, but on how prior executions are distilled, related, and recomposed.

To address this gap, we propose \textbf{EpiDroid}, a black-box, pluggable framework for semantic state dependency-aware mobile GUI exploration. EpiDroid distills raw execution traces into stable test fragments and leverages an LLM to perform semantic reasoning over these fragments and infer their underlying dependencies. It then applies a \emph{Recomposition-Replay} paradigm to identify high-value mutable state elements and deterministically replay affected branches. Through iterative feedback, the framework refines a state-dependency graph and uses it to drive existing testing tools toward previously unreachable deep states. The name \textbf{EpiDroid} is inspired by the Greek mythological figure \emph{Epimetheus}, whose name literally means \emph{hindsight}. This reflects the central design of our approach: rather than relying only on forward expansion, EpiDroid looks back over executed branches, analyzes what they changed, and replays dependent branches under new state conditions.

In summary, this paper makes the following contributions:
\begin{itemize}
    \item \textbf{Recomposition-Replay paradigm.} We propose a new GUI testing perspective inferring cross-path state dependencies and replaying recomposed sequences to reach deep states inaccessible via traditional forward exploration.
    \item \textbf{The EpiDroid framework.} We propose a black-box, pluggable framework that equips automated GUI testing tools with LLM-driven state awareness and dependency inference.
    \item \textbf{Comprehensive evaluation.} We conduct comprehensive evaluations by integrating EpiDroid into both industrial and state-of-the-art research testing tools across 20 real-world apps. Results show that EpiDroid consistently improves all baselines, raising average code coverage by 10--28\% and delivering 3--4$\times$ more coverage gain than continuing the baseline alone from the same starting point.
\end{itemize}
\section{Motivation and Problem Statement}
\label{sec:motivation}

\begin{figure}[t]
    \centering
    \includegraphics[width=\columnwidth]{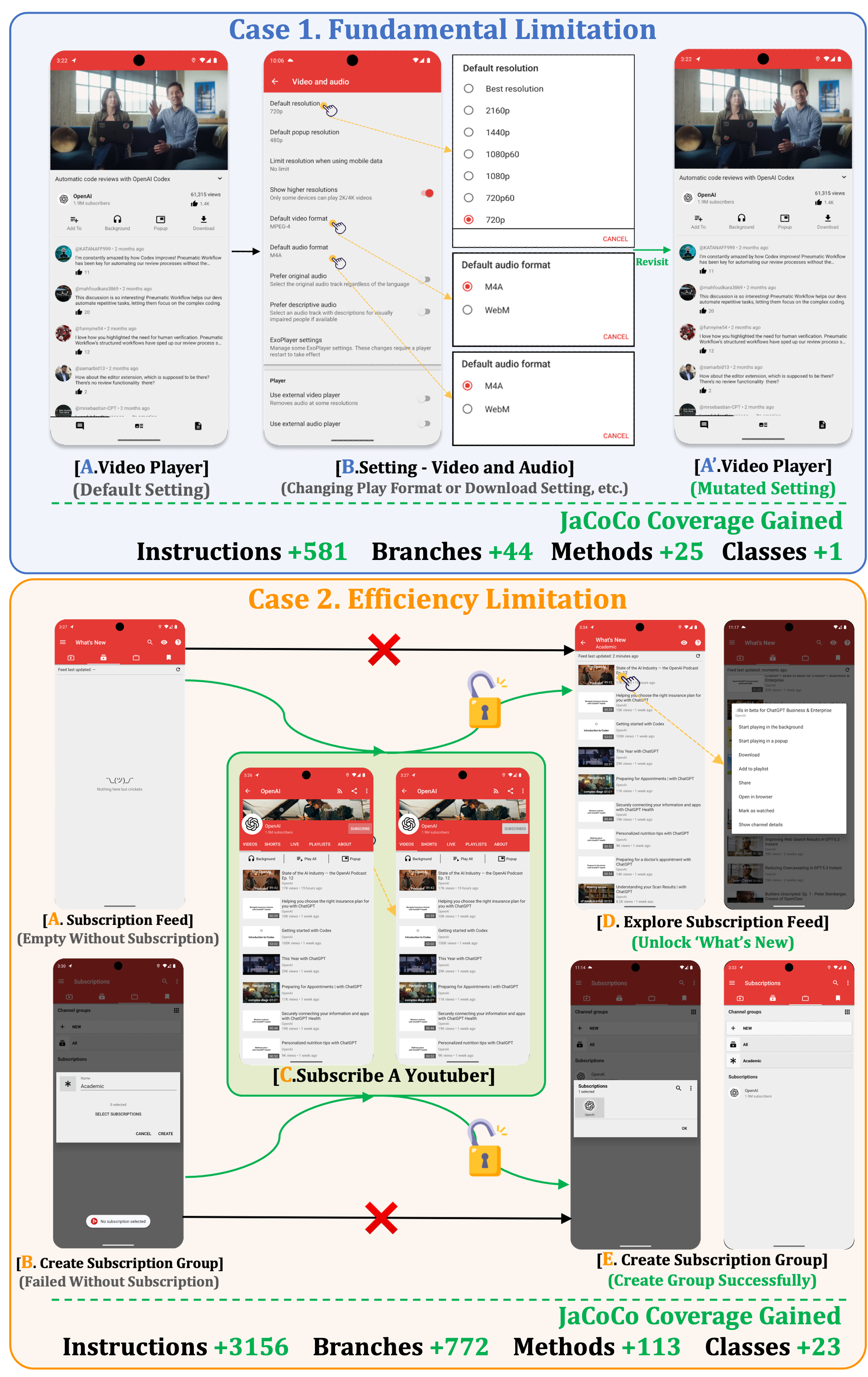}
    \caption{Motivation cases in NewPipe. (Case 1) shows a structural limitation where a state-dependent branch is unreachable via forward exploration; (Case 2) shows an efficiency limitation where preconditions are not satisfied before exploring dependent branches.}
    \label{fig:motivation}
\end{figure}

To investigate how the lack of state and dependency awareness limits existing exploration techniques, we conducted an instrumentation study on NewPipe, an open-source media application. Using JaCoCo\cite{JaCoCo} to identify uncovered branches, we analyzed the execution results of two state-of-the-art LLM-enhanced tools (Figure~\ref{fig:motivation}).

LLM-Explorer uses an LLM to maintain knowledge and guide exploration by selecting unexplored components on the current page. In contrast, LLMDroid applies LLM guidance primarily when coverage stalls, favoring functional paths that transition to new pages. Despite these different strategies, both tools show systematic failures when handling behaviors that depend on specific states.

\textbf{Case 1. Fundamental Limitation.} The upper half of Figure~\ref{fig:motivation} illustrates a state dependency where the target behavior is triggered only through a non-monotonic sequence: Branch~A $\rightarrow$ Branch~B $\rightarrow$ Branch~A. In NewPipe, Branch~A opens a video player with the default media settings. Branch~B enters the settings panel where the user can change the default resolution, audio format, or download configuration---modifying a global mutable state. Re-executing Branch~A after this mutation loads the player under the updated settings, unlocking previously unreachable code paths (Instructions~+581, Branches~+44). In our experiments, both tools failed to generate this sequence. LLM-Explorer marked Branch~A as already explored after the first visit and only interacted with new components on the current page. Similarly, LLMDroid prioritized transitions to new activities; since re-entering the player lacked immediate navigational novelty, it was consistently deprioritized. This indicates that without explicit modeling of mutable states and state-conditioned replay, configuration-dependent behaviors remain unreachable.

\textbf{Case 2. Efficiency Limitation.} The lower half of Figure~\ref{fig:motivation} shows that even when target behaviors are theoretically reachable, they are often missed within a limited execution budget. Branch~A (viewing the subscription feed) and Branch~B (creating a subscription group) both fail initially because no creators have been followed. Branch~C (subscribing to a YouTuber) establishes the necessary precondition; once satisfied, re-visiting Branch~A unlocks the updated feed (Branch~D), and re-visiting Branch~B successfully creates a group (Branch~E), together yielding Instructions~+3156 and Branches~+772. Current methods prioritize immediately reachable or locally novel targets, wasting resources on A and B before the precondition in C is met. Even if C is eventually executed, the remaining budget rarely suffices for the required re-execution of A or B. Consequently, exploration remains biased toward shallow, configuration-independent features. These cases suggest that improving coverage requires both explicit UI state modeling and intentional action sequencing to unlock hidden application logic.

\subsection{Problem Statement}

The cases above share a common structural pattern: a target behavior becomes reachable only after a separate interaction path mutates the application's internal state. We now formalize this observation.

Given a target application $P$, an automated testing tool generates a set of execution traces $\mathcal{T} = \{\tau_1, \tau_2, \dots\}$ within a time budget $B$, where each trace $\tau$ is a sequence of GUI states and events with per-step coverage:
$$\tau = s_{0} \xrightarrow{e_{1}, C_{1}} s_{1} \xrightarrow{e_{2}, C_{2}} \cdots \xrightarrow{e_{n}, C_{n}} s_{n}$$
Here $s_i$ is an abstract GUI state, $e_i$ the executed event, and $C_i$ the code or activity coverage gained at that step. The exploration goal is to maximize the cumulative coverage $C(\mathcal{T})$ within $B$. In practice, raw traces are noisy and contain non-deterministic transitions caused by asynchronous rendering, network latency, or animations. A trace is \emph{replayable} if repeating its action sequence from the same initial state consistently yields the same state trajectory and coverage increments; raw traces must be stabilized into such replayable fragments before any cross-path reasoning can be applied.

We call a UI component whose interactions persistently alter application state a \emph{Mutable State Element} (MSE), formally defined as a triple $m = \langle \sigma_{sea}, \Delta\Sigma, \text{Scope} \rangle$: $\sigma_{sea}$ is the state-effecting action sequence (e.g., toggling a switch then confirming), $\Delta\Sigma$ the observed mutation evidence (e.g., a flipped \texttt{is\_checked} attribute), and $\text{Scope} \in \{Inter\text{-}page, Global, Intra\text{-}page\}$ the estimated impact scope. The mutation fragment $g$ in a cross-path dependency is precisely the sub-trace that exercises an MSE to establish a precondition $\phi$.

We define a \emph{cross-path state dependency} as follows. Let $f$ be a target fragment that reaches deep application behavior, and let $\phi$ be a precondition on the application's internal state that must hold for $f$ to trigger new coverage. If $\phi$ is established by a mutation fragment $g$ on a different execution path, then the composite sequence $g \circ f$, executing $g$ to establish $\phi$ then replaying $f$ under the mutated state, constitutes a cross-path state dependency. In Case~1, $g$ corresponds to the settings change (Branch~B) and $f$ to the player re-entry (Branch~A); in Case~2, $g$ is the subscription action (Branch~C) and $f$ is the feed or group access (Branch~A/B).

Existing automated testing tools operate under a forward-execution paradigm and face two inherent limitations against such dependencies:

\begin{enumerate}
    \item \textbf{No re-visitation mechanism.} Once a path $f$ has been explored, forward testing tools deprioritize re-execution. Even if a subsequent mutation $g$ establishes the precondition $\phi$, the tool has no mechanism to recognize that $f$ should be re-executed under the new state (Case~1).
    \item \textbf{No dependency-aware scheduling.} GUI exploration tools cannot reason about which mutations establish useful preconditions for other paths. Without such knowledge, they waste budget on paths whose preconditions are unsatisfied and may exhaust the remaining budget before the required re-execution occurs (Case~2).
\end{enumerate}

These limitations are not artifacts of specific tool implementations but are inherent to the forward-execution paradigm. Overcoming them requires explicitly (1)~identifying mutable state elements and their cross-path impact, (2)~inferring which fragments depend on which state mutations, and (3)~recomposing and replaying dependent fragments under mutated conditions. EpiDroid is designed to address all three requirements as a plugin layer over existing GUI testing tools.
\section{Approach}
\label{sec:approach}

\subsection{EpiDroid Overview}

EpiDroid follows a three-stage workflow---trace stabilization and state-change tracking, semantic profiling and impact reasoning, and dependency-guided sequence recomposition---and is deployed as a plug-in layer above existing GUI testing tools. Using the trace and MSE abstractions defined in Section~\ref{sec:motivation}, the system first distills raw traces into replayable fragments with explicit mutation evidence, then uses the LLM to construct semantic dependencies and prioritize high-value MSEs, and finally executes dependency-guided composite replay to reach deep states.
\begin{figure*}
    \centering
    \includegraphics[width=\textwidth]{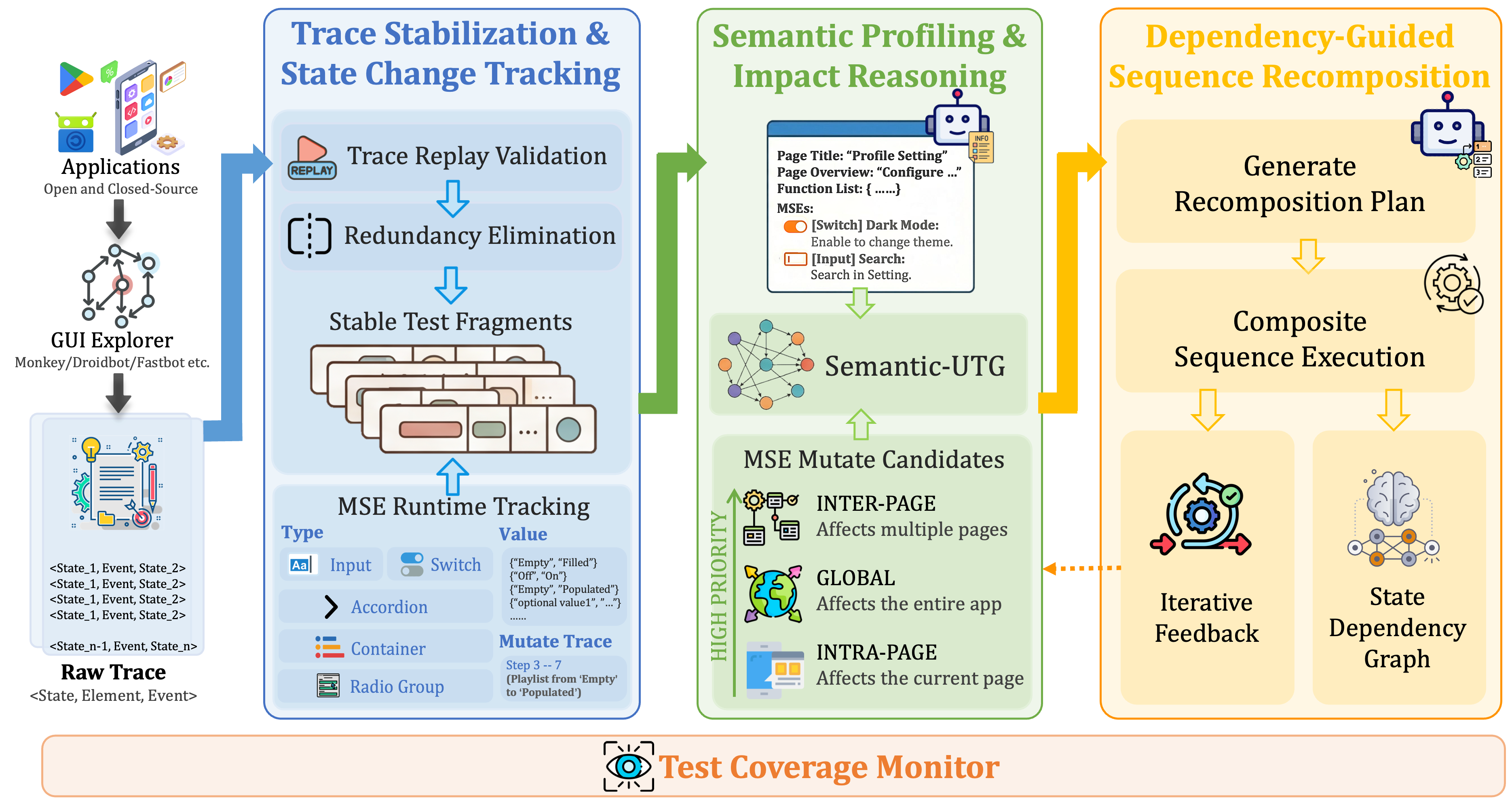}
    \caption{Framework overview of EpiDroid. Stage~1 stabilizes raw traces and extracts replayable fragments with runtime mutation evidence. Stage~2 performs semantic profiling and impact reasoning to infer cross-page dependencies and prioritize MSEs. Stage~3 synthesizes and replays dependency-guided composite sequences, and feeds execution outcomes back to refine the semantic dependency graph iteratively.}
    \label{fig:overview}
    \Description{Framework overview diagram showing EpiDroid's three-stage pipeline. Stage 1 takes raw exploration traces and produces stabilized replayable fragments with mutation records. Stage 2 performs LLM-based semantic profiling to infer page summaries, validate MSEs, and analyze impact scopes. Stage 3 executes dependency-guided composite sequences through navigation, mutation, and dependent replay, with feedback signals looping back to refine the semantic dependency graph.}
\end{figure*}

EpiDroid is built upon DroidBot~\cite{Li17b} and inherits its UI transition graph (UTG) modeling infrastructure. As shown in Figure~\ref{fig:overview}, the system comprises three stages connected through explicit artifacts. Stage~1 produces minimized, replayable test fragments and MSE tracking records, while Stage~2 constructs a cluster-level Semantic-UTG via LLM and outputs prioritized, validated MSEs. Stage~3 then uses these artifacts for dependency-guided recomposition. Feedback from this final stage, including confirmed dependencies, priority adjustments, and newly discovered MSEs, is used to refine the Semantic-UTG. New MSEs further trigger additional rounds of Stage~2 profiling. Throughout this process, a cross-stage coverage monitor tracks code and activity coverage to guide prioritization and evaluate incremental gains. This design decouples high-speed physical execution from semantic reasoning, allowing EpiDroid to improve state discovery without replacing the underlying testing tool.

\subsection{Trace Stabilization and State-Change Tracking}

This stage refines raw exploration traces by eliminating non- deterministic noise and redundant events, producing validated, replayable test fragments with explicit tracking of state changes.

\subsubsection{Trace Slicing and Redundancy Reduction}

EpiDroid groups GUI states into \emph{page clusters} based on the structural similarity of their content-free abstract view trees (measured by Dice coefficient), so that states with similar layout structures but differing in dynamic content are treated as equivalent. Building upon the DetReduce algorithm~\cite{Cho18}, EpiDroid refines raw traces through trace slicing, replay-based stabilization, and redundancy elimination.
Subsequently, each obtained sequence undergoes replay verification to ensure reproducibility. Under the default verification strategy, each sequence is replayed once. A sequence is considered successfully reproduced and marked as stable if the final state after replay belongs to the expected page cluster observed during the original execution.
When replay fails due to non-deterministic behavior, EpiDroid initiates a fallback mechanism to restore the execution context. Specifically, the system first attempts to navigate to the target state using a shortest-path strategy within the explored interface state graph. If this strategy is not feasible, the system skips the event that caused the failure and proceeds with the remaining operations. If neither recovery strategy successfully restores the execution flow, the sequence is truncated to retain only the prefix that was successfully replayed.
Following the replay process, the system further eliminates redundant operations from the validated sequences. Specifically, EpiDroid removes redundant loops and tail events. Redundant loops refer to sub-traces that start and end at the same abstract state without contributing new coverage. Tail events are operations at the end of a sequence that do not produce new state transitions or coverage. The removal of these redundant structures results in a set of more compact and stable execution fragments while maintaining the coverage of the original behavior.

\subsubsection{Runtime Mutable State Elements Tracking}
During the automated exploration process, EpiDroid performs runtime tracking of mutable state elements (MSEs) to capture the trajectory of state changes during interactions. The framework monitors the GUI component tree and identifies potential MSEs by detecting changes in state-bearing attributes (e.g., \texttt{checked}, \texttt{selected}, \texttt{text}) before and after each interaction. Specifically, it recognizes five component types, where the first four admit a binary state ab\-strac\-tion and the last supports multiple values:
\begin{itemize}[nosep,leftmargin=2em]
    \item \textbf{Switch/Toggle}: On/Off or Checked/Unchecked;
    \item \textbf{Input}: Empty/Filled;
    \item \textbf{Expandable}: Expanded/Collapsed;
    \item \textbf{Container}: Empty/Populated;
    \item \textbf{Radio Group}: 2--4 mutually exclusive options (e.g., language or quality settings).
\end{itemize}
For each MSE candidate, EpiDroid records its component identifier, the page cluster it belongs to, the set of observed values ($\Delta\Sigma$ evidence), and the action sequence that triggered the change ($\sigma_{sea}$). These structured MSE tracking records provide the empirical basis for the semantic validation and impact reasoning performed in Stage~2.

\subsection{Semantic Profiling and Impact Reasoning}
The second stage abstracts execution traces into a high-level semantic representation of application functionality. By using the contextual reasoning of the LLM, EpiDroid transforms stabilized trace fragments and observed attribute changes into a semantically enhanced UTG while identifying and prioritizing Mutable State Elements (MSEs) for subsequent exploration.
\subsubsection{Semantic State Validation}
Although the first stage identifies potential MSEs through heuristic rules, physical attribute changes, such as layout displacements caused by animations or temporary text replacements, often introduce significant non-functional noise. To ensure precise exploration, EpiDroid utilizes the LLM to perform semantic validation.

\textbf{Functional Summarization}: For each page cluster $C$ established in Stage~1, EpiDroid provides a denoised GUI component tree~\cite{Vu24} and the associated interaction history sequence as context to the LLM. The LLM then generates a functional summary of the cluster.This process integrates business semantics into traditional structural abstractions.
\textbf{MSE Validation}: The LLM validates the initially identified MSEs using the annotated view tree, where each candidate component is augmented with its observed value history and unvisited states derived from $\Delta\Sigma$. It determines whether a physical change, such as toggling a switch, represents a genuine state mutation that affects the application's business logic rather than a transient UI fluctuation.

\textbf{Filtering Criteria}: The LLM filters out non-functional fluctuations, such as the appearance or disappearance of loading animations and minor UI style adjustments. This ensures that subsequent exploration focuses exclusively on components that substantially alter the execution environment or the data context of the application.

\subsubsection{Impact Scope Analysis and Prioritization}
After validating the effectiveness of the MSEs, EpiDroid further reasons about the associated impact scope. We formally define the impact scope as a set $\mathcal{L} = \{Global, Inter\text{-}page, Intra\text{-}page\}$ and utilize the LLM to classify each MSE based on the business logic of the application.
\textbf{Inter-page} impact involves mutations that modify the behavior, data, or functional availability of other specific page clusters. For example, clicking ``Add to Cart'' on a ``Detail Page'' results in a state change on the ``Checkout Page.'' Inter-page dependencies often serve as prerequisite thresholds for unlocking deep functionalities and are therefore assigned the highest exploration priority. \textbf{Global} impact refers to mutations that modify the core execution environment or global configurations, such as changing the media decoding format or switching the language. Although they may affect all explored pages, global mutations risk an explosion of the exploration space and are ranked second. \textbf{Intra-page} impact refers to mutations whose influence is restricted to the current page cluster, such as filtering local search results or expanding/collapsing a UI panel.

EpiDroid prioritizes inter-page MSEs for recomposition because they directly represent the cross-path dependencies where mutations on one page unlock behavior on another. The system maintains validated MSEs in a priority queue ranked primarily by scope, following the order of inter-page, global, and intra-page. Within each scope, MSEs with unvisited values take precedence over those already fully observed, ensuring that the recomposition budget focuses on mutations most likely to reveal new behaviors. The outputs of this stage, including functional summaries, annotated MSEs, and the prioritized mutation queue, are combined with the STG from Stage~1 to form a cluster-level Semantic-UTG ($G_{sem}$), which serves as the shared knowledge base for Stage~3.

\subsection{Dependency-Guided Sequence Recomposition}
The third stage involves constructing and executing composite test cases based on the high-priority Mutable State Elements (MSEs) identified previously. By reasoning about the impact of mutation operations, EpiDroid generates non-monotonic execution paths or sequences with specific dependency orderings that traditional forward exploration tools often miss due to a lack of state awareness.

\subsubsection{Mutation Impact Reasoning}
The system first selects high-value MSEs from the priority queue as mutation targets and follows a structured logical reasoning process. During the \textbf{Mutation Execution} phase, it adopts an ``unvisited value first'' strategy to select specific values for a component. For instance, if the WebM format has not been tested previously, the system sets the state target of the MSE as $\Sigma_{target} = \{format: "WebM"\}$. In the \textbf{Impact Inference} phase, the LLM receives the current mutation target and its semantic context—including detailed information of the page cluster $C_{mut}$ where the mutation occurs and all reachable page clusters—to infer which page clusters and functional components may exhibit behavioral shifts. Finally, for \textbf{Composite Plan Generation}, the LLM produces a logical scheme that non-linearly combines the mutation sequence $\sigma_{sea}$ with the affected path fragments requiring re-validation, rather than generating a simple linear sequence. This process addresses the structural limitations described in Case 1 by answering the critical question: ``After changing state $S$, which historical page should be revisited to observe the consequences?''

\subsubsection{Composite Sequence Execution}
This step transforms abstract composite schemes into practical executable test sequences by concatenating verified test fragments. The execution workflow consists of three primary phases. First, during the \textbf{Navigation} phase, the system utilizes the simplified UTG generated in the first stage to compute and execute the shortest path from the current state to the target mutation page $C_{mut}$. Second, in the \textbf{State Mutation} phase, the system executes the designated mutation sequence $\sigma_{sea}$. For MSEs whose state transitions require multi-step interactions---such as populating a Container by filling a form and confirming a dialog---the system invokes the LLM to guide the completion of the mutation, generating the necessary intermediate actions based on the current page context. Third, in the \textbf{Dependent Replay} phase, the system dynamically appends fragments of the affected paths to the post-mutation state. By replaying these fragments under the new configuration $\Sigma_{new}$, the system monitors whether previously explored branches trigger any new states. When a previously unseen page state is encountered during replay, the system performs a bounded BFS exploration over the reachable action space of the new state to maximize immediate coverage gain.

\begin{algorithm}[t]
\caption{Composite Sequence Execution}
\label{alg:recomposition}
\begin{algorithmic}[1]
\Require MSE $m = \langle \sigma_{sea}, \Delta\Sigma, \text{Scope} \rangle$, mutation page $C_{mut}$, target value $\Sigma_{target}$, affected pages $\mathcal{P}_{aff}$, UTG $G$
\Ensure Execution feedback $\mathcal{F}$
\State $s_{cur} \leftarrow$ current state
\State $S_{new} \leftarrow \emptyset$, $\Delta C \leftarrow 0$
\State $C_{cur} \leftarrow \textsc{Cluster}(s_{cur})$
\State $\pi \leftarrow \textsc{ShortestPath}(G, C_{cur}, C_{mut})$ \Comment{Navigate to mutation page}
\For{each event $e \in \pi$}
    \State Execute $e$; update $s_{cur}$
\EndFor
\State Execute $\sigma_{sea}$ on $s_{cur}$ toward $\Sigma_{target}$ \Comment{Perform mutation}
\If{$\Delta\Sigma$ not observed}
    \Return $\mathcal{F} \leftarrow \textsc{OperationalFailure}$
\EndIf
\For{each $C_{aff} \in \mathcal{P}_{aff}$}
    \State Navigate to $C_{aff}$
    \State Replay verified fragments of $C_{aff}$ under $\Sigma_{target}$
    \State $S_{new} \leftarrow S_{new} \cup \textsc{NewStates}()$; $\Delta C \leftarrow \Delta C + \textsc{CovGain}()$
    \If{$\textsc{NewStates}() \neq \emptyset$}
        \State $S_{new} \leftarrow S_{new} \cup \textsc{BFS}(s_{cur})$; $\Delta C \leftarrow \Delta C + \textsc{CovGain}()$
    \EndIf
\EndFor
\If{$|S_{new}| > 0$ or $\Delta C > 0$}
    \Return $\mathcal{F} \leftarrow \textsc{PositiveDiscovery}(S_{new}, \Delta C)$
\Else
    \Return $\mathcal{F} \leftarrow \textsc{SemanticMismatch}$
\EndIf
\end{algorithmic}
\end{algorithm}

\subsubsection{Iterative Feedback and State Discovery}
EpiDroid treats each recomposed execution as a "probing process" to refine the state model of the application. The execution results are categorized into three feedback signals to drive the iterative evolution of the semantic state-aware UTG ($G_{sem}$). \textbf{Positive Discovery} occurs if the recomposed sequence successfully triggers a new abstract state $s_{new}$ or increases the cumulative branch coverage $C(T)$; the system confirms the predicted dependency and persists it in $G_{sem}$, establishing a verified functional association between the mutation and the new behavior. \textbf{Semantic Mismatch} happens if a sequence is replayable but fails to produce any new state increments, indicating a semantic offset between the actual results and the predictions of the LLM. In such cases, the feedback mechanism decreases the priority of the relevant MSE to ensure that the exploration budget is not wasted on "false positive" impact scopes. \textbf{Operational Failure} is identified when a sequence cannot be replayed correctly due to non-deterministic factors. Following the trajectory reduction strategy, the system records the shortest non-replayable prefix for path pruning in the next iteration.

A single iteration concludes once all candidate MSEs have been processed through the plan--execute--feedback cycle. If the execution discovers new MSEs that were not present in the original candidate set, a new iteration is triggered: the newly identified MSEs are fed back to Stage~2 for semantic profiling and impact reasoning, and the resulting mutation points enter Stage~3 for the next round of recomposition. The process terminates when no new MSEs are discovered during an iteration. Through this closed-loop process of ``mutation--validation--feedback,'' EpiDroid progressively completes the state dependency graph of the application, achieving a paradigm shift from ``blind probing'' to ``goal-driven exploration.''

\section{Evaluation}
\label{sec:evaluation}

This section presents our evaluation of EpiDroid. The experimental subjects and setup are described in Section~\ref{subsec:setup}. The evaluation addresses three research questions:

\begin{itemize}
\item \textbf{RQ1:} How effective is EpiDroid when integrated with existing Android GUI explorers?
\item \textbf{RQ2:} How do trace stabilization and dependency-guided reasoning contribute to EpiDroid's effectiveness?
\item \textbf{RQ3:} What are the LLM cost and marginal coverage return of EpiDroid?
\end{itemize}

\subsection{Experimental Setup}
\label{subsec:setup}

\noindent\textbf{Benchmark Applications.}
To evaluate EpiDroid as a semantic enhancement layer, we assembled a benchmark of 20 popular Android applications from the \textbf{Google Play Store} and \textbf{F-Droid}. These applications span diverse categories including finance, social networking, system tools, productivity, and media players, and exhibit a wide range of complexity as reflected by their method and activity counts at the code level and by the functional diversity of their UI pages. To ensure that results are not biased by source-code accessibility, we maintained an equal 1:1 split between open-source (F-Droid) and closed-source (Google Play Store) applications. All benchmark APKs are provided in both arm64 and x86\_64 builds to support heterogeneous emulator configurations. We intentionally retain apps with low absolute coverage (e.g., OceanEx, Fing) to stress-test EpiDroid under unfavorable conditions such as authentication walls and hardware-gated functionality, ensuring that our average results are not inflated by cherry-picking only favorable subjects. Table~\ref{tab:dataset} summarizes the benchmark.

\begin{table}[t]
    \centering
    \caption{Overview of the evaluated Android applications.}
    \label{tab:dataset}
    \small
    \resizebox{\columnwidth}{!}{
    \begin{tabular}{l l c c}
    \toprule
    \textbf{App Name} & \textbf{Category} & \textbf{Activities} & \textbf{Complexity} \\
    \midrule
    \multicolumn{4}{l}{\textit{Open-Source (F-Droid)}} \\
    \midrule
    AntennaPod, AP & Music & 11 & Low \\
    ScarletNotes, SN & Productivity & 8 & Low \\
    Drawing Pad, DP & Art \& Design & 15 & Low \\
    OpenTracks, OT & Health & 23 & Low \\
    NewPipe, NP & Multimedia & 12 & Medium \\
    AnkiDroid, AD & Education & 26 & Medium \\
    Tasks, TK & Productivity & 41 & Medium \\
    FirefoxLite, FL & Communication & 27 & Medium \\
    My Expenses, ME & Finance & 51 & Medium \\
    Wikipedia, WK & Books & 58 & High \\
    \midrule
    \multicolumn{4}{l}{\textit{Closed-Source (Google Play Store)}} \\
    \midrule
    Word Web, WW & Books & 11 & Low \\
    Daily Workouts, DW & Health & 31 & Low \\
    Sunrise \& Sunset, SS & Weather & 3 & Low \\
    AccuBattery, AB & Tools & 13 & Low \\
    Time Planner, TP & Productivity & 6 & Medium \\
    HabitNow, HN & Productivity & 30 & Medium \\
    BusinessCalendar2, BC & Productivity & 41 & Medium \\
    OceanEx, OE & Finance & 81 & High \\
    Fing, FG & Network & 92 & High \\
    SmartNews, SM & News & 108 & High \\
    \bottomrule
    \end{tabular}
    }
\end{table}

\noindent\textbf{Baselines.}
EpiDroid was compared against two classes of baseline explorers. The first class comprises traditional explorers: \textit{Monkey}~\cite{Pat18}, the most widely adopted random-based tool, and \textit{Fastbot}~\cite{Lv22}, a widely deployed industrial-strength explorer that represented the strongest pre-LLM baseline. The second class consists of two state-of-the-art LLM-based methods: \textit{LLMDroid}~\cite{Wan25} and \textit{LLM-Explorer}~\cite{Zha25}, both of which leverage LLM to guide exploration.

\noindent\textbf{Metrics.}
Several metrics were employed. \textit{Average Code Coverage} ($ACC$) measures the method-level code coverage across all benchmark applications. \textit{Average Activity Coverage} ($AAC$) captures the proportion of declared activities visited during exploration. \textit{Time-To-Coverage} ($TTC$) denotes the wall-clock time required to reach the maximum coverage that the corresponding baseline achieves in its full run, measuring how quickly EpiDroid can match or surpass baseline performance. \textit{Replay Success Rate} ($RSR$) indicates the percentage of recomposed action sequences that execute successfully on real devices.

\noindent\textbf{Environment.}
All experiments were conducted on four identically configured Android~10.0 emulators with a Pixel~8~Pro profile (API~29). All LLM-dependent components (including EpiDroid and the LLM-based baselines) uniformly employ \textbf{GPT-5 mini} as the backbone model. Each configuration is repeated three times, and we report the average to mitigate randomness in both exploration and LLM responses. Code coverage is collected via \textbf{Androlog}~\cite{Sam24}, a black-box instrumentation tool that performs bytecode-level rewriting on compiled APKs, enabling uniform method-level coverage measurement for both open-source and closed-source applications. Based on it, we implemented a unified coverage monitoring module across all tools.

\subsection{RQ1: Overall Effectiveness in State Discovery}

This research question evaluates the coverage gain of EpiDroid when integrated with existing explorers. The experimental protocol adopts a two-phase design with a total budget of 1~hour. In the warm-up phase, each baseline explorer runs for 30~minutes to produce initial traces $\tau_{\mathit{init}}$. In the enhancement phase, either the baseline continues for another 30~minutes (\textit{Baseline-Ext}), or EpiDroid enhances the collected traces for 30~minutes (\textit{Baseline+Epi}). This split is motivated by prior studies~\cite{Cho15,Su17} and corroborated by recent LLM-based tools~\cite{Wan25,Zha25}, which show that automated exploration typically saturates within approximately 30~minutes. Our own progression data (Table~\ref{tab:rq3_progression}) corroborates this: continuing the baseline beyond 30~minutes yields only +1.08--1.53\,pp additional $ACC$, confirming that forward exploration has largely saturated by that point.

We integrate EpiDroid with Fastbot~(FB), LLM-Explorer~(LE), and LLMDroid~(LD), yielding FB+Epi\-Droid, LE+Epi\-Droid, and LD+Epi\-Droid. Monkey is included as a coverage reference but not as an integration target, because its purely random event stream does not produce structured traces amenable to fragment extraction.

\begin{table*}[t]
    \centering
    \caption{Code coverage ($ACC$, \%) per application. $^\dag$Open-source. Green/red values show relative change. FB=Fastbot, LE=LLM-Explorer, LD=LLMDroid.}
    \label{tab:rq1_acc}
    \footnotesize
    \renewcommand{\arraystretch}{0.92}
    \resizebox{\textwidth}{!}{
    \begin{tabular}{l r r r r r r r}
    \toprule
    \textbf{App} & \textbf{Monkey} & \textbf{Fastbot} & \textbf{FB+EpiDroid} & \textbf{LLM-Explorer} & \textbf{LE+EpiDroid} & \textbf{LLMDroid} & \textbf{LD+EpiDroid} \\
    \midrule
    AP$^\dag$ & 15.59 & 36.55 & 40.32 \gain{+10.32} & 40.85 & \textbf{44.88} \gain{+9.87} & 44.08 & 44.67 \gain{+1.34} \\
    SN$^\dag$ & 8.53 & 12.99 & 15.66 \gain{+20.60} & 11.75 & \textbf{20.86} \gain{+77.44} & 13.98 & 19.66 \gain{+40.59} \\
    DP$^\dag$ & 40.25 & 47.10 & 51.67 \gain{+9.69} & 41.83 & 51.57 \gain{+23.27} & 50.44 & \textbf{53.63} \gain{+6.33} \\
    OT$^\dag$ & 7.82 & 8.71 & 9.41 \gain{+8.07} & 9.36 & \textbf{11.29} \gain{+20.60} & 9.06 & 10.91 \gain{+20.53} \\
    NP$^\dag$ & 23.24 & 13.98 & 19.91 \gain{+42.45} & 31.49 & 34.23 \gain{+8.68} & 25.51 & \textbf{36.47} \gain{+42.97} \\
    AD$^\dag$ & 5.42 & 7.70 & 8.31 \gain{+7.89} & 6.03 & 9.81 \gain{+62.74} & 7.89 & \textbf{9.84} \gain{+24.78} \\
    TK$^\dag$ & 24.65 & 43.00 & 47.31 \gain{+10.02} & 26.52 & \textbf{55.46} \gain{+109.14} & 45.46 & 49.88 \gain{+9.72} \\
    FL$^\dag$ & 25.17 & 42.13 & 41.70 \loss{-1.03} & 29.11 & 32.13 \gain{+10.37} & 42.09 & \textbf{42.38} \gain{+0.69} \\
    ME$^\dag$ & 13.56 & 15.45 & 17.77 \gain{+15.03} & 16.31 & 18.31 \gain{+12.28} & 17.68 & \textbf{18.99} \gain{+7.45} \\
    WK$^\dag$ & 28.80 & 31.24 & 33.45 \gain{+7.08} & 31.05 & 34.23 \gain{+10.24} & 35.58 & \textbf{37.24} \gain{+4.69} \\
    WW & 24.48 & \textbf{28.81} & 27.99 \loss{-2.85} & 24.10 & 27.34 \gain{+13.45} & 28.19 & 28.55 \gain{+1.29} \\
    DW & 22.36 & 19.05 & 19.91 \gain{+4.53} & 22.90 & 22.36 \loss{-2.38} & \textbf{24.95} & 24.88 \loss{-0.29} \\
    SS & 8.63 & 8.51 & 8.93 \gain{+4.93} & 7.80 & 9.09 \gain{+16.45} & 8.74 & \textbf{10.23} \gain{+17.06} \\
    AB & 24.57 & 22.30 & 25.79 \gain{+15.64} & 13.86 & 15.23 \gain{+9.92} & \textbf{27.45} & 26.77 \loss{-2.48} \\
    TP & 27.00 & 30.10 & 34.50 \gain{+14.61} & 16.50 & 32.85 \gain{+99.08} & 32.47 & \textbf{36.54} \gain{+12.56} \\
    HN & 32.42 & 33.65 & 38.43 \gain{+14.19} & 27.77 & 34.38 \gain{+23.81} & 38.02 & \textbf{39.35} \gain{+3.52} \\
    BC & 19.12 & 20.21 & 22.27 \gain{+10.21} & 18.94 & 23.08 \gain{+21.82} & \textbf{26.51} & 26.27 \loss{-0.92} \\
    OE & 12.54 & 12.33 & \textbf{13.33} \gain{+8.13} & 11.26 & 12.86 \gain{+14.24} & 13.25 & 13.23 \loss{-0.15} \\
    FG & 21.03 & 28.78 & \textbf{28.79} \gain{+0.05} & 23.41 & 25.87 \gain{+10.49} & 28.73 & 28.77 \gain{+0.14} \\
    SM & 12.45 & 19.44 & 20.30 \gain{+4.43} & 18.14 & \textbf{20.54} \gain{+13.25} & 14.74 & 19.24 \gain{+30.52} \\
    \midrule
    \textbf{AVG.} & \textbf{19.88} & \textbf{24.10} & \textbf{26.29} \gain{+10.20} & \textbf{21.45} & \textbf{26.82} \gain{+28.24} & \textbf{26.74} & \textbf{28.88} \gain{+11.02} \\
    \bottomrule
    \end{tabular}
    }
\end{table*}

Table~\ref{tab:rq1_acc} reports per-app code coverage. Within the same 1-hour budget, all three enhanced configurations outperform their baselines in average $ACC$. FB+ EpiDroid improves over Fastbot by 10.20\%, LE+EpiDroid over LLM-Explorer by 28.24\%, and LD+EpiDroid over LLMDroid by 11.02\%.

\noindent\textbf{Interpretation of the improvement gap.}
The relative improvement of LE+EpiDroid (28.24\%) is notably higher than that of FB+ EpiDroid (10.20\%) and LD+EpiDroid (11.02\%). As Table~\ref{tab:events} shows, this disparity stems from the underlying engines' event throughput. DroidBot, on which LLM-Explorer is built, must construct a UI Transition Graph during exploration, pausing after each event to capture the UI hierarchy and update the graph, which limits it to 227~events per 30-minute session. Fastbot, which underlies both Fastbot and LLMDroid, uses a native on-device injection mechanism without such per-step modeling overhead, achieving 499 and 440~events respectively. The lower throughput of DroidBot results in lower warm-up coverage for LE, leaving a larger margin for EpiDroid to improve upon. Conversely, LD already combines Fastbot's high throughput with LLM guidance, pushing its baseline coverage closer to a natural ceiling rather than reflecting a limitation of EpiDroid.

\begin{table}[t]
    \centering
    \caption{Average GUI events per application (over 20~apps).}
    \label{tab:events}
    \small
    \renewcommand{\arraystretch}{0.92}
    \begin{tabular*}{\columnwidth}{l @{\extracolsep{\fill}} r r r}
    \toprule
    & \textbf{Fastbot (FB)} & \textbf{LLMDroid (LD)} & \textbf{LLM-Explorer (LE)} \\
    \midrule
    30\,min & 499 & 440 & 227 \\
    60\,min & 962 & 875 & 469 \\
    \bottomrule
    \end{tabular*}
\end{table}

The largest absolute gains appear on applications with abundant mutable state elements (MSEs) and multi-level state dependencies: Tasks (+109.14\%), Time~Planner (+99.08\%), and ScarletNotes (+77.44\%), all under LE+EpiDroid. In these applications, EpiDroid's iterative recomposition resolved sequential prerequisites that 1-hour baseline runs consistently missed, e.g., creating a task list before editing subtasks in Tasks, or configuring account settings before creating events in Time~Planner.

\noindent\textbf{Negative cases.}
A small number of regressions are observed. Word Web (FB+EpiDroid: $-$2.85\%) and Daily~Workouts (LE+EpiDroid: $-$2.38\%) exhibit marginal losses due to their flat, shallow page structures that offer limited room for recomposition. FirefoxLite shows a slight drop under FB+EpiDroid ($-$1.03\%) because WebView-rendered content produces UI hierarchies partially opaque to semantic abstraction. Fing (FB+EpiDroid: +0.05\%, LD+EpiDroid: +0.14\%) shows near-zero gain despite 92~activities: its core functionality relies on network scanning and hardware probing, making most transitions unreachable in an emulator. Similarly, OceanEx stays at roughly 12--13\% coverage across all configurations, as its authentication wall blocks the majority of states.

Overall, these results indicate that EpiDroid's advantage is most pronounced in applications with deep, stateful navigation and abundant MSEs, while apps with simple page structures, heavy WebView reliance, or environment-gated functionality benefit less from semantic enhancement.

\noindent\textbf{Applicability Boundaries.}
The effectiveness of EpiDroid is contingent on the accessibility of UI hierarchies through standard Android accessibility APIs. Applications that heavily rely on WebView-rendered content or custom rendering engines present partial opacity to semantic abstraction, as the framework cannot inspect dynamically generated DOM structures within embedded web contexts. Similarly, apps with server-enforced authentication walls or hardware-dependent functionality (network scanning, sensor probing) inherently limit the reachable state space in emulated environments, regardless of the exploration strategy employed. More broadly, EpiDroid's MSE tracking relies on GUI-level observability: state mutations that do not manifest as visible attribute changes in the view hierarchy (e.g., background database writes or silent network requests) cannot be captured by the current heuristic. These represent structural constraints of black-box testing rather than limitations specific to EpiDroid's recomposition mechanism. Future work may address these scenarios through lightweight instrumentation or integration with accessibility metadata enrichment.

Table~\ref{tab:rq1_summary} further breaks down the average $ACC$ and $AAC$ by source type. EpiDroid consistently improves code coverage on both open-source and closed-source applications, with open-source apps benefiting more (LD+EpiDroid $ACC$: 32.37\% vs.\ 25.38\%). This gap is expected: open-source apps in our benchmark tend to have richer UI logic and more MSEs exposed through standard Android widgets, whereas several closed-source apps rely on custom rendering or server-gated features that limit recomposition opportunities.

For $AAC$, LD+EpiDroid improves over LLMDroid across both subsets (36.24\% vs.\ 34.11\% overall), as navigation during the recomposition phase occasionally reaches previously unvisited activities. In contrast, LE+EpiDroid shows lower $AAC$ than its baseline (20.00\% vs.\ 24.28\%). This reflects a deliberate design trade-off: EpiDroid replaces the second 30~minutes of broad forward exploration with targeted recomposition replay, which deepens code coverage within already-visited activities rather than expanding the activity frontier. Since method-level code coverage ($ACC$) is the primary metric for assessing functional adequacy---reaching more activities is only valuable if it also exercises new code paths---this trade-off is favorable in practice. When activity breadth is also important, a hybrid strategy that alternates recomposition with short bursts of forward exploration could mitigate the $AAC$ reduction; we leave this to future work.

\begin{table}[t]
    \centering
    \caption{Average coverage grouped by source type. $ACC$\,=\,code coverage; $AAC$\,=\,activity coverage.}
    \label{tab:rq1_summary}
    \footnotesize
    \renewcommand{\arraystretch}{0.95}
    \resizebox{\columnwidth}{!}{
    \begin{tabular}{l rrr rrr}
    \toprule
    & \multicolumn{3}{c}{\textbf{ACC (\%)}} & \multicolumn{3}{c}{\textbf{AAC (\%)}} \\
    \cmidrule(lr){2-4} \cmidrule(lr){5-7}
    \textbf{Tool} & \textbf{Open} & \textbf{Closed} & \textbf{All} & \textbf{Open} & \textbf{Closed} & \textbf{All} \\
    \midrule
    Monkey        & 19.30 & 20.46 & 19.88 & 22.35 & 23.83 & 23.09 \\
    Fastbot       & 25.89 & 22.32 & 24.10 & 28.19 & 25.96 & 27.08 \\
    FB+EpiDroid   & 28.55 & 24.02 & 26.29 & 30.75 & 27.18 & 28.96 \\
    \midrule
    LLM-Explorer  & 24.43 & 18.47 & 21.45 & 26.17 & 22.39 & 24.28 \\
    LE+EpiDroid   & 31.28 & 22.36 & 26.82 & 21.14 & 18.85 & 20.00 \\
    \midrule
    LLMDroid      & 29.18 & 24.31 & 26.74 & 35.99 & 32.22 & 34.11 \\
    LD+EpiDroid   & \textbf{32.37} & \textbf{25.38} & \textbf{28.88} & \textbf{39.19} & \textbf{33.30} & \textbf{36.24} \\
    \bottomrule
    \end{tabular}
    }
\end{table}

\subsection{RQ2: Ablation Study}

To isolate each component's contribution, we conduct ablation experiments using warm-up traces from LLMDroid, the strongest baseline on average, so that improvements are measured against a high-quality starting point.

\begin{itemize}
    \item \textbf{EpiDroid}: the complete framework.
    \item \textbf{EpiDroid\textsubscript{NoS}}: skips Stage~1 trace stabilization; raw traces are passed directly to the reasoning and recomposition stages.
    \item \textbf{EpiDroid\textsubscript{NoR}}: skips Stage~2; MSEs are selected randomly and composed without dependency inference.
\end{itemize}

Table~\ref{tab:rq2_ablation} presents the ablation results. Both components are critical, but affect different aspects of the pipeline.

\begin{table}[t]
    \centering
    \caption{Ablation study of EpiDroid (averaged over 20~apps). RSR=Replay Success Rate; Eff.\ Iters=iterations with at least one new state.}
    \label{tab:rq2_ablation}
    \small
    \renewcommand{\arraystretch}{0.95}
    \begin{tabular*}{\columnwidth}{l @{\extracolsep{\fill}} r r r}
    \toprule
    \textbf{Variant} & \textbf{ACC (\%)} & \textbf{RSR (\%)} & \textbf{Eff. Iters} \\
    \midrule
    EpiDroid                        & 28.88 & 75 & 3.4 \\
    EpiDroid\textsubscript{NoS}     & 26.04 & 38 & 1.2 \\
    EpiDroid\textsubscript{NoR}     & 26.91 & 79 & 2.5 \\
    \bottomrule
    \end{tabular*}
\end{table}

\noindent\textbf{Trace Stabilization Analysis.}
A core premise of EpiDroid is that raw warm-up traces are too noisy and redundant to serve directly as recomposition inputs. Stage~1 addresses this by clustering GUI events into page-level groups, identifying unique transitions, and pruning redundant steps. Using LLMDroid's warm-up traces, the average raw trace contains 440~steps across 29~page clusters with 67~unique transitions. The average redundancy ratio is 63.4\% (including replay failures and steps with no coverage gain), meaning nearly two-thirds of recorded events do not contribute new coverage. After stabilization, traces are reduced to roughly 278~active steps on average. Apps with few distinct pages but long sequences (e.g., Daily~Workouts: 96.5\% redundancy, ScarletNotes: 93\%) see the largest pruning, while apps with many unique transitions retain more steps. This compression is critical for the subsequent recomposition cycle: without it, the replay engine would spend its budget re-executing redundant sequences, and the LLM reasoning module would receive noisy, repetitive page descriptions. As Table~\ref{tab:rq2_ablation} confirms, removing stabilization (EpiDroid\textsubscript{NoS}) causes RSR to drop from 75\% to 38\%, and effective iterations fall from 3.4 to 1.2, meaning EpiDroid essentially degrades to a single-round attempt before convergence.

\noindent\textbf{Dependency Reasoning Analysis.}
EpiDroid\textsubscript{NoR} retains trace stabilization but removes LLM-based dependency inference. Instead of performing mutation impact reasoning and targeted replay, it randomly selects a stabilized fragment for replay. Because these fragments are already validated by Stage~1, replay success rate remains high (79\%, slightly above the full pipeline's 75\%). However, ACC drops by 1.97\,pp (28.88\% $\to$ 26.91\%) and effective iterations fall from 3.4 to 2.5. This gap reveals that raw replay of arbitrary fragments rarely discovers new states: the coverage gain in the full pipeline primarily comes from deliberately mutating high-value MSEs and then replaying semantically related branches under the changed conditions. Without dependency reasoning to identify which MSEs to target and which branches to revisit, most replays simply re-traverse already-covered paths. This confirms that stabilization ensures replay reliability, while dependency reasoning is responsible for directing the limited budget toward state changes that actually unlock new functionality.

\subsection{RQ3: Efficiency and Practicality}

This research question evaluates the practical deployment profile of EpiDroid, specifically focusing on the LLM cost associated with the enhancement phase and the marginal coverage return relative to a continuation of the baseline.

\noindent\textbf{Cost and Marginal Coverage Return.}
Table~\ref{tab:rq3_progression} presents a direct comparison of the second 30-minute interval under two distinct strategies: continuing the baseline versus transitioning to EpiDroid, both initiated from an identical warm-up state. The design of EpiDroid incorporates a \textit{sparse calling} mechanism; specifically, the LLM is invoked solely for page-level semantic summarization and mutation reasoning rather than for every GUI action. This approach leads to fewer invocations (34--41 versus 26--31) but yields higher token consumption per call (${\sim}$2,500 tokens versus ${\sim}$1,800). Such an increase is attributable to the fact that each call includes comprehensive page cluster descriptions, MSE mutation histories, and dependency graphs.\footnote{Cost estimated based on GPT-5 mini pricing (\$0.25/M input, \$2.00/M output) at the time of evaluation.}

Despite a comparable cost structure (\$0.09--\$0.10 versus \$0.05--\$0.08 per application), EpiDroid achieves a substantially higher gain in coverage. A continuation of LLMDroid adds only +1.08\,pp to the $ACC$; in contrast, replacing the second half of the testing period with EpiDroid yields an increase of +3.22\,pp (\textbf{3.0$\times$}). This disparity is even more pronounced for LLM-Explorer: +1.53\,pp for the continuation versus +6.30\,pp with EpiDroid (\textbf{4.1$\times$}). These results demonstrate that dependency-guided recomposition represents a more cost-effective utilization of the latter half of the testing budget than sustained forward exploration.

\begin{table*}[t]
    \centering
    \caption{Cost and coverage comparison of the second 30-minute phase (averaged over 20~apps). Left: LLM cost of the second phase only. Right: code coverage ($ACC$) at each stage; $\Delta$ is the gain over the shared warm-up. LE=LLM-Explorer, LD=LLMDroid.}
    \label{tab:rq3_progression}
    \small
    \renewcommand{\arraystretch}{0.95}
    \begin{tabular*}{\textwidth}{@{\extracolsep{\fill}}l rrrr rrr@{}}
    \toprule
    & \multicolumn{4}{c}{\textbf{LLM Cost (2nd 30\,min)}} & \multicolumn{3}{c}{\textbf{Code Coverage $ACC$ (\%)}} \\
    \cmidrule(lr){2-5} \cmidrule(lr){6-8}
    \textbf{Strategy} & \textbf{Calls} & \textbf{Input (K)} & \textbf{Output (K)} & \textbf{Cost (\$)} & \textbf{Warm-up} & \textbf{Final} & \textbf{$\Delta$ (pp)} \\
    \midrule
    LD continued     & 31 & 45 & 21 & 0.05 & 25.66 & 26.74 & +1.08 \\
    LD+EpiDroid      & 41 & 61 & 42 & 0.10 & 25.66 & 28.88 & \textbf{+3.22} \textit{(3.0$\times$)} \\
    \midrule
    LE continued     & 26 & 25 & 35 & 0.08 & 19.08 & 20.61 & +1.53 \\
    LE+EpiDroid      & 34 & 49 & 38 & 0.09 & 19.08 & 25.38 & \textbf{+6.30} \textit{(4.1$\times$)} \\
    \bottomrule
    \end{tabular*}
\end{table*}

\section{Related Work}
\label{sec:related}

\subsection{Traditional Automated GUI Testing}

Traditional Android GUI testing primarily focuses on scalable event generation and broad state-space traversal. Early tools like Dynodroid \cite{Mac13} improved random exploration with system-aware event generation, while Sapienz \cite{Mao16} optimized coverage and crash discovery through multi-objective search. DroidBot \cite{Li17b} demonstrated lightweight UI-guided exploration using UI Transition Graphs. These systems established strong baselines for automated GUI testing, but their strengths lie mainly in breadth-oriented exploration rather than systematic access to deep state. Subsequent work improved these baselines through better models and stronger learning signals. Stoat \cite{Su17} and APE \cite{Gu19} used model-based exploration with stochastic guidance or model abstraction and refinement to improve effectiveness. Humanoid \cite{Li19}, Q-testing \cite{Pan20}, and Fastbot2 \cite{Lv22} introduced deep learning and reinforcement learning to prioritize actions exposing new functionality. These methods improved traversal quality but remained centered on advancing the current frontier, generally failing to model how combining semantically related fragments from different paths could reconstruct the latent preconditions of deep app behavior.

Several studies expose parts of the problem targeted by EpiDroid: Gesda~\cite{Guo20} shows that dependency knowledge improves coverage and bug finding; ComboDroid~\cite{Wan20} demonstrates that complex functionality often requires combinations of shorter use cases; Fax~\cite{Yan20} reveals that distant activities need richer launching contexts; and TimeMachine~\cite{Don20} and Sara~\cite{Guo19} show that replaying from discovered states is more effective than rebuilding prefixes from scratch. These results collectively motivate dependency-aware recomposition, yet none provides a black-box framework that infers semantic dependencies over reusable fragments and exploits them systematically. EpiDroid builds on these insights but differs in two key aspects: it requires no static analysis or source-code access, operating entirely as a black-box plugin; and it unifies dependency inference, fragment composition, and state replay into a single closed-loop framework rather than addressing each in isolation.

\subsection{LLM-Based Mobile GUI Exploration}

The rise of LLMs has introduced stronger semantic reasoning into mobile GUI testing~\cite{Arn18,Sai20,Li23,Yu23}. Early explorations such as DroidBot-GPT~\cite{Wen23b} and AdbGPT~\cite{Liu23c} demonstrated the feasibility of using chatbot-style LLMs to automate Android interactions. GPTDroid~\cite{Liu23} frames testing as an interactive question-answering process and uses functionality-aware memory to guide exploration. DroidAgent~\cite{Yoo24} moves further toward semantic automation by generating and pursuing realistic intent-driven tasks. Both systems show that language models can improve interaction quality beyond what is possible with purely heuristic or learned action selection.

A newer wave of work focuses on practicality. LLMDroid~\cite{Wan25} augments existing testing tools with selective LLM guidance once autonomous exploration begins to plateau. LLM-Explorer~\cite{Zha25} reduces token cost further by using LLMs mainly for compact knowledge maintenance rather than step-wise control. MemoDroid~\cite{Che25} introduces dynamic memory so that exploration experience can be reused across runs and apps. TestWeaver~\cite{Wan25b} reduces repeated reasoning cost by reusing LLM decisions for shared UI interactions across test cases. Vision-based approaches such as VisionDroid~\cite{Liu24} leverage multimodal LLMs to detect non-crash functional bugs through visual understanding. ProphetAgent~\cite{Kon25} synthesizes GUI tests from automated property extraction, and GUIPilot~\cite{Liu25c} introduces consistency-based exploration for more systematic coverage. LLM-guided scenario-based testing~\cite{Yu25} and LELANTE~\cite{Sha25} further diversify the range of LLM-integrated strategies. More recently, CovAgent~\cite{Min26} addresses hard-to-reach functionality by reasoning about hidden activation conditions and synthesizing dynamic instrumentation for blocked activities.

These methods advance semantic guidance, cost control, and memory reuse, but their dominant abstractions remain local: next-action reasoning, task pursuit, prompt memory, or selective invocation. Even CovAgent~\cite{Min26}, which directly targets hard reachability, relies on code inspection and instrumentation rather than black-box recomposition. In contrast, EpiDroid treats cross-path state dependency as an explicit, refinable object and operates as a plugin layer that requires no source-code access, complementing rather than replacing any of the above explorers.
\section{Conclusion and Future Work}

We present EpiDroid, a black-box, pluggable framework that augments existing mobile GUI testing tools with semantic state dependency awareness. By distilling execution traces into stable fragments, inferring cross-path dependencies, and applying a Recomposition Replay paradigm, EpiDroid systematically unlocks deep application states that forward exploration alone cannot reach. Experimental results show that EpiDroid consistently and substantially improves both industrial and state-of-the-art research tools, while its sparse-calling design keeps LLM cost comparable to existing baselines. 

EpiDroid opens a new research direction for enhancing automated mobile testing through dependency-aware trace recomposition. Based on this framework, a diverse range of techniques can be explored in future work. Promising directions include element-level dependency inference for finer-grained state mutation capture, convergence-aware adaptive stopping for more efficient budget allocation, richer context signals (e.g., network responses, accessibility metadata) for handling WebView and authentication-gated scenarios, and generalization to cross-application and multi-device settings where state dependencies span process boundaries and demand scalable abstraction strategies.

\clearpage
\bibliographystyle{ACM-Reference-Format}
\bibliography{refs}










\end{document}